\documentclass{sf2a-conf2010}
\usepackage{graphicx}
\usepackage{hyperref}
\usepackage[]{natbib}

\begin{document}
\TitreGlobal{SF2A 2010}
%
\title{Stratospheric Observatory for Infrared Astronomy}
\author{M. Hamidouche}\address{USRA - NASA Ames Research Center, MS N211-3, Moffett Field, CA 94035}
\author{E. Young$^1$}
\author{P. Marcum}\address{NASA Ames Research Center, MS 211-1, Moffett Field, CA 94035}
\author{A. Krabbe}\address{Deutsches SOFIA Institut, Pfaffenwaldring 31, 70569 Stuttgart, Germany}

\runningtitle{SOFIA}
%
\setcounter{page}{237}
\index{Hamidouche, M.}
\index{Young, E.}
\index{Marcum, P.}
\index{Krabbe, A.}

\maketitle
\begin{abstract}
We present one of the new generations of observatories, the Stratospheric Observatory For Infrared Astronomy (SOFIA). This is an airborne observatory consisting of a 2.7-m telescope mounted on a modified Boeing B747-SP airplane. Flying at an up to 45,000 ft (14 km) altitude, SOFIA will observe above more than 99 percent of the Earth's atmospheric water vapor allowing observations in the normally obscured far-infrared. We outline the observatory capabilities and goals. The first-generation science instruments flying on board SOFIA and their main astronomical goals are also presented.

%
%
\end{abstract}
\section{Introduction}

SOFIA (Stratospheric Observatory For Infrared Astronomy) consists of a 2.7-meter telescope mounted in a modified Boeing 747‐SP aircraft. SOFIA is a joint project of NASA and the Deutsches Zentrum f\"ur Luft‐ und Raumfahrt (DLR). Operations costs and observing time will be shared by the United States (80\%) and Germany (20\%). It is a near-space observatory that comes home after every flight. Flying at altitudes up to 45,000 ft (14 km), SOFIA observes from above more than 99\% of Earth's atmospheric water vapor. SOFIA will begin science observations in 2011. It will offer the international astronomical community approximately 1000 science observing hours per year for two decades, when full operational capabilities are reached in 2014. Science proposals will be open to the international community. In this paper, we focus on SOFIA scientific capabilities. 

\section{SOFIA Performance}

The first generation science instruments are being tested or under development by different institutions in both the US and Germany, including imaging cameras and spectrographs as well as imaging cameras with spectrometers. SOFIA will observe at wavelengths from 0.3 $\mu$m up to 1600 $\mu$m. It will be capable of high resolution spectroscopy (R $>$ 10$^4$) at wavelengths between 5 and 240 $\mu$m (Figure 1). The 8 arcminute diameter field of view will ultimately allow use of very large format detector arrays. SOFIA will provide diffraction limited imaging long-ward of 15 $\mu$m. After Herschel's cryogen depletion, SOFIA will be the only telescope covering the 30 to 300 $\mu$m wavelength range in the next years.

\section{SOFIA Uniqueness for Astronomy}

One of its great strengths is that its scientific instruments can be exchanged regularly to
accommodate changing science requirements and new technologies. Furthermore, large, massive, complex and sophisticated instruments with substantial power and heat dissipation needs can be
flown on SOFIA, and thus increasing SOFIA’s science productivity.

SOFIA has unique capabilities for studying transient events. The observatory can
operate from air-bases worldwide to respond to new discoveries
in both the northern and southern hemispheres. SOFIA has the flexibility to
respond to events such as supernovae and nova explosions, cometary impacts,
comet apparitions, eclipses, occultations, near Earth objects, activity in Active Galactic Nuclei, and activity in luminous variable stars.
SOFIA's wide range of instruments will facilitate a coordinated science program
through analysis of specific targets. No other observatory operating in SOFIA’s
wavelength range can provide such a large variety of available instruments for such a long period of time. A particular advantage of SOFIA is that it will be able
to access events unavailable to many space observatories because of the viewing
constraints imposed by their orbits. For example, SOFIA can observe astrophysical events which occur closer to the Sun than most spacecrafts can. This will enable temporal monitoring of
supernovae, novae, and variable stars throughout the year. SOFIA’s 20-year operational lifetime will enable long term temporal studies and
follow‐up of work initiated by SOFIA itself and by other observatories. Many
space missions are relatively short compared with the critical cycle of observation,
analysis, and further observation. The Herschel observatory will raise scientific questions that will benefit
from follow‐up observations well after their missions have ended. SOFIA will keep the community engaged in fundamental science research until the next generation of missions is launched.

\bigskip
\bigskip
\bigskip
\bigskip



\begin{figure}[ht!]
 \centering
 \includegraphics[width=1.\textwidth]{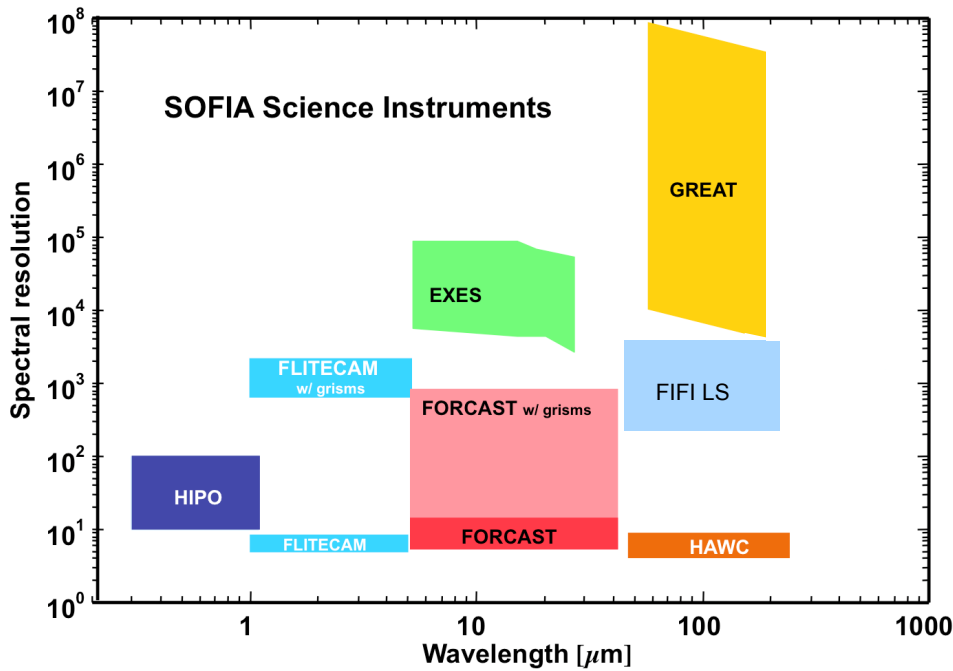}      
  \caption{SOFIA first generation instruments shown in a plot of spectral resolution vs. the wavelength.}
  \label{fig1}
\end{figure}

\end{document}